# Gamma Rossi-alpha, Feynman-alpha and Gamma Differential Die-Away concepts as a potential alternative/complement to the traditional thermal neutron based analysis in Safeguards

Dina Chernikova[a,*], Syed F. Naeem[a], Nermin Trnjanin[a], Kåre Axell[a,b], Anders Nordlund[a]

[a]*Chalmers University of Technology, Department of Applied Physics, Nuclear Engineering, Fysikgården 4, SE-412 96 Göteborg, Sweden*
[b]*Swedish Radiation Safety Authority, SE-171 16 Stockholm, Sweden*

**Abstract**

A new concept for thermal neutron based correlation and multiplicity measurements is proposed in this paper. The main idea of the concept consists of using 2.223 MeV gammas (or 1.201 MeV, DE) originating in the $^1H(n,\gamma)^2D$-reaction instead of using traditional thermal neutron counting. Results of investigations presented in this paper indicate that gammas from thermal neutron capture reaction preserve the information about the correlation characteristics of thermal (fast) neutrons in the same time scale. Therefore, instead of thermal neutron detectors (or as a complement) one may use traditional and inexpensive gamma detectors, such NaI, BGO, CdZnTe or any other gamma detectors. In this work we used D8x8 cm$^2$ NaI scintillator to test the concept. Thus, the new approach helps to address the problem of replacement of $^3$He-counters and problems related to the specific measurements of spent nuclear fuel directly in the spent fuel pool. It has a particular importance for nuclear safeguards and security. Overall, this work represents the proof of concept study and reports on the experimental and numerical evidence that thermal neutron capture gammas may be used in the context of correlation and multiplicity measurements. Investigations were performed for a $^{252}$Cf-correlated neutron source and an $^{241}$Am-Be-random neutron source. The related idea of Gamma Differential Die-Away approach is investigated numerically in this paper as well, and will be tested experimentally in a future work.

*Corresponding author, email: dina@nephy.chalmers.se

## 1. Introduction

Thermal neutron-based detectors, such as $^3$He gas proportional detectors, have been around already for decades and are used in a variety of Nuclear Safeguards and Security applications, portal monitors, coincidence counters, waste assay systems, etc. Therefore, the present problem of the global shortage of $^3$He [1] (the total $^3$He demand is ~65 $10^3$ l/y, while the total supply is ~10-20 $10^3$ l/y [2]) led to the number of international programs initiated to find a possible replacement for $^3$He detection technology. A number of different new detectors and systems were suggested as a replacement, a good overview of each of them can be found in a recent report of [3]. In this work we approach the problem from another perspective and investigate the possibility of instead using of thermal neutrons to use 2.223 MeV gammas which are produced during the thermalization process by the by epithermal and thermal neutrons in hydrogen-containing material. This depends mostly on the yield of 2.223 MeV gammas (ratio between thermal neutrons and capture gammas) and the ability of 2.223 MeV gammas to carry the same time-related information as thermal neutrons. If 2.223 MeV gammas perform satisfactory in these two aspects then the traditional and inexpensive gamma detectors can be used as a potential alternative to the $^3$He thermal neutron detectors for correlation measurements and counting.

## 2. The main concept

The main idea of the concept consists of using 2.223 MeV gammas or 1.201 MeV gammas (Double Escape (DE)) originating in the $^1H(n,\gamma)^2D$-reaction instead of using traditional thermal neutron counting. The high-energy neutrons from the fissionable material or spent nuclear fuel get slowed down to ephithermal and thermal energies in the hydrogen-containing material (water, polyethylene etc.). A part of these thermal neutrons undergo $^1H(n,\gamma)^2D$-reaction with the release of 2.223 MeV gammas which are finally detected by the gamma detector, as shown in Figure 1. It is important to notice that out of all emitted neutrons (for the case of $^{252}$Cf in the 40x40x30 cm water tank, as shown in Figure 2) approximately half undergo $^1H(n,\gamma)^2D$ reaction after thermalization. Results of investigations indicate that these gammas carry the same information about the correlations as the thermal neutrons. Moreover, due to the origin of the gammas, since it takes on average microseconds for a neutron to be slowed down and captured, the time scale stays the same as the one for thermal neutrons. As our previous research [4, 5] show, the source gamma will be correlated on the ns scale which makes their analysis somewhat cumbersome. Therefore, the beauty of using a 2.223 MeV gamma line is partly related to the microseconds time scale since there will be no interference with other lines in the gamma spectrum which will be correlated on the same microsecond time scale, unless another strong neutron absorber is present in the system which can lead to the production of the highly energetic gamma lines. Though, even in this case the overall information will be improved.

## 3. Description of the simulation process and the experimental set-up

The new concept was tested experimentally and numerically. The experiments were performed with a weak $^{252}$Cf (~17.3 kBq) spontaneous fission source of correlated neutrons and an $^{241}$Am-Be ($\alpha$ ,n)-source of random neutrons with a neutron yield of approximately $1.1 \cdot 10^7$ neutrons per second. Due to the lack of access to the real spent fuel assembly and a portable neutron generator at the moment, the potential of the method for a spent nuclear fuel measurements directly in the spent fuel pool and the related idea of Gamma Differential Die-Away approach were investigated only numerically in this paper. Future experimental work on these issues is foreseen.

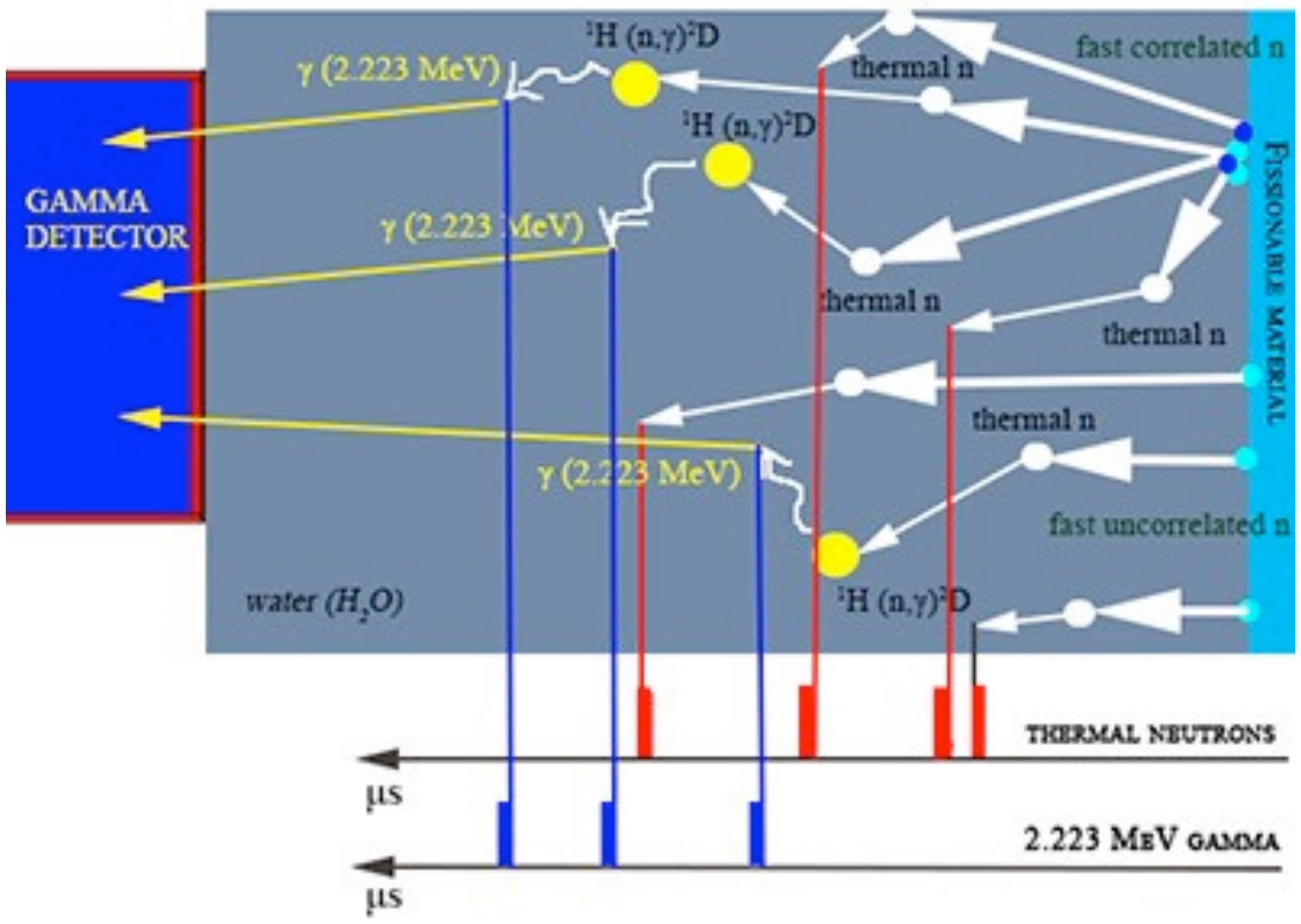

Figure 1: A simplified illustration of the proposed concept.

### *3.1. Geometry of the experimental set-up*

A Plexiglas water tank was constructed specifically for experimental investigations, as shown in Figure 2.

Dimensions of the tank was 40x40x30 cm$^3$. The tank was equipped with water tight Plexiglas sample (source) holder to avoid direct contact of the sample and the water in the tank and provide the possibility to place the sample directly in the center of the tank. One NaI scintillator (D8 x 8 cm$^2$) was used in the measurements.

Two sources were used for the calibration procedure to cover the energy range from 511 keV to 2.5 MeV, i.e. a $^{22}$Na (<1 MBq) source related to the 0.511 MeV, 1.274 MeV lines and a coincident line of 1.785 MeV; a $^{60}$Co source with lines of 1.17 MeV, 1.33 MeV and a coincident line of 2.5 MeV.

### *3.2. Electronics and settings used in experiments*

The NaI detector was connected to a 8 channel, 12 bit 250 MS/s, VX1720E CAEN digitizer. The high voltage bias was adjusted for the detector using a CAEN (Mod. SY403) 64 channel high voltage supply system as 1365.4 V.

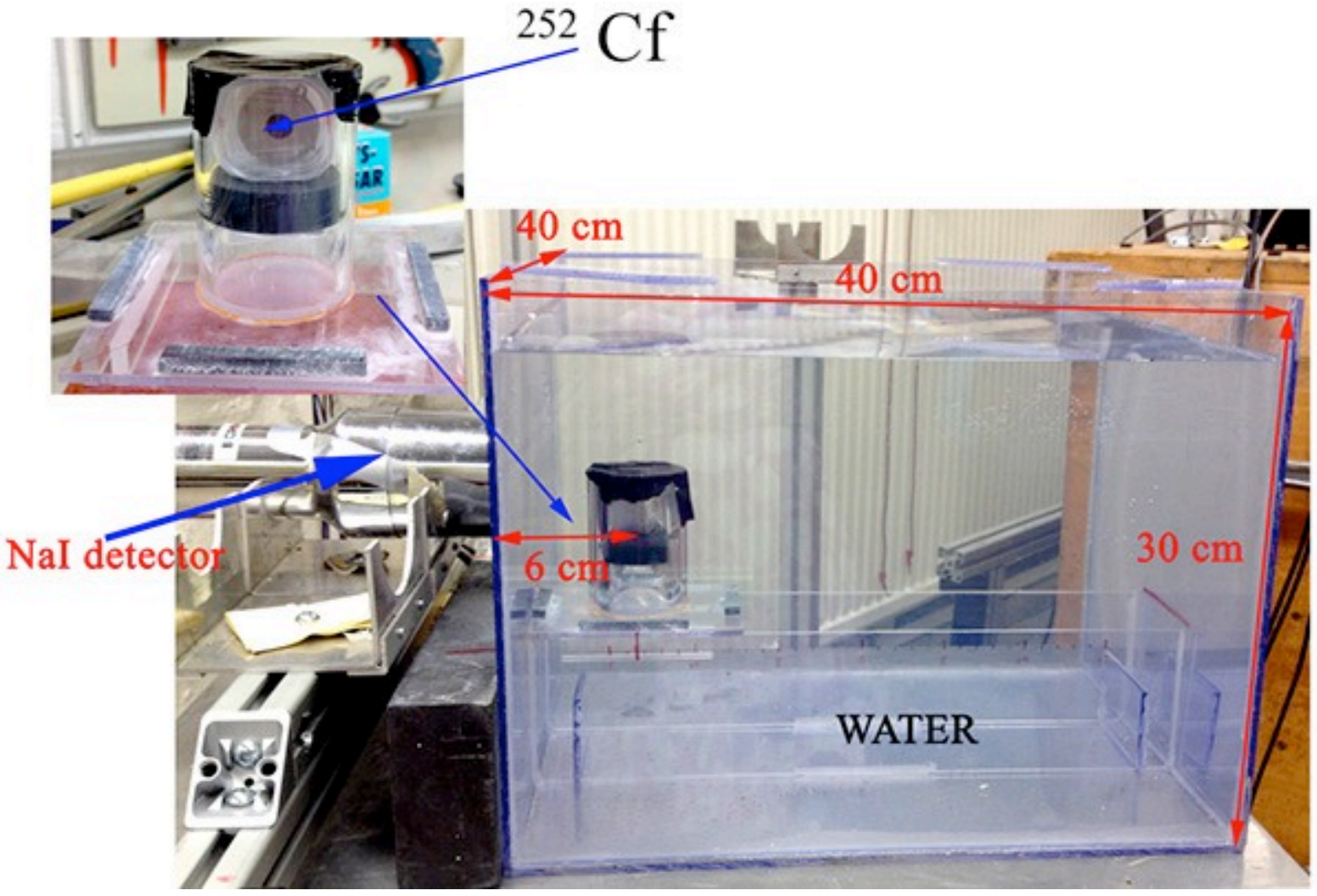

Figure 2: A simplified illustration of the proposed concept.

### *3.3. Data processing procedure*

The experimental evaluation of the Rossi-alpha distribution and Feynman-alpha (variance to mean) ratio for thermal neutron-induced gammas was performed with one detector in a traditional way [6, 7], as shown in Figure 3.

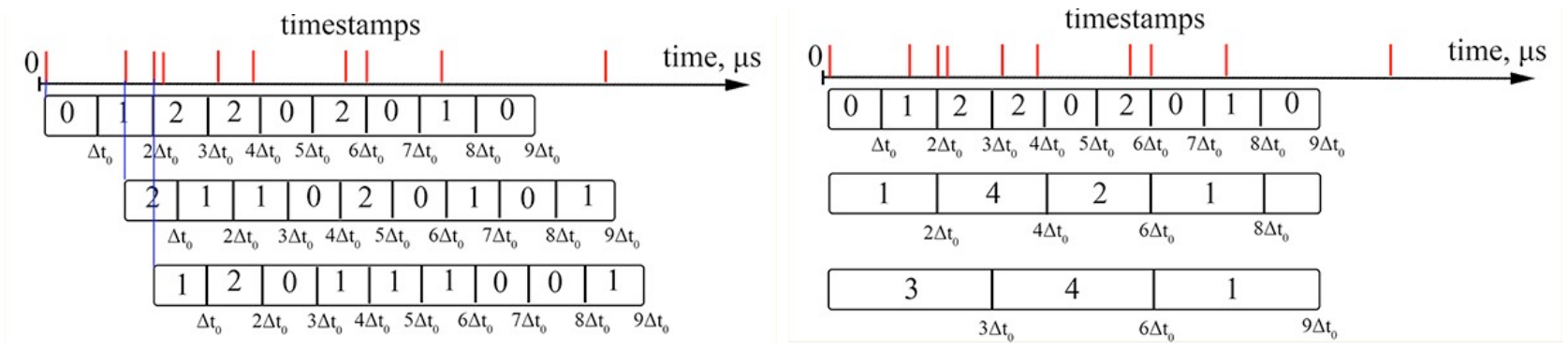

Figure 3: Procedures of data processing for traditional Rossi-alpha (l.h.s) and Feynman-alpha (r.h.s.) measurements.

Originally, the digitized waveform and corresponding timestamps were collected during the interval of time which varied from 30 minutes to two hours. Each individual pulse was post processed offline so the eventual pulse train included timestamps and the pulse heights of the signals. In the Feynman-alpha measurement the variance and mean of the numbers of counts (N) was evaluated in k consecutive time intervals of length $\Delta t_0$ (10 $\mu s$).

The uncertainty in the variance to mean ratio was estimated as below [**?** ]:

---

In the Rossi-alpha measurement, the autocovariance of the numbers of counts (N) was evaluated in short time intervals of length $\Delta t_0$ (8 $\mu s$) after each detector count, as shown in l.h.s. of Figure 3.

In the case of only random events being detected the variance to mean ratio will represent unity while the Rossi-alpha distribution will be constant with time.

### *3.4. Geometry and details of the simulation set-up*

In the numerical studies the experimental geometry was modeled as a 40x40x30 $cm^3$ water tank with a source and cylindrical detector. To check the applicability of the concept for the investigations of a spent fuel assembly directly in the fuel pool (water filled) numerical studies were performed with a SVEA-64 assembly which contained five gadolinium-bearing poison rods, one water rod and the fuel rods in four sub-assemblies separated by a water moderator (water cross). The model and design dimensions of the assembly are shown in Figure 4 [8].

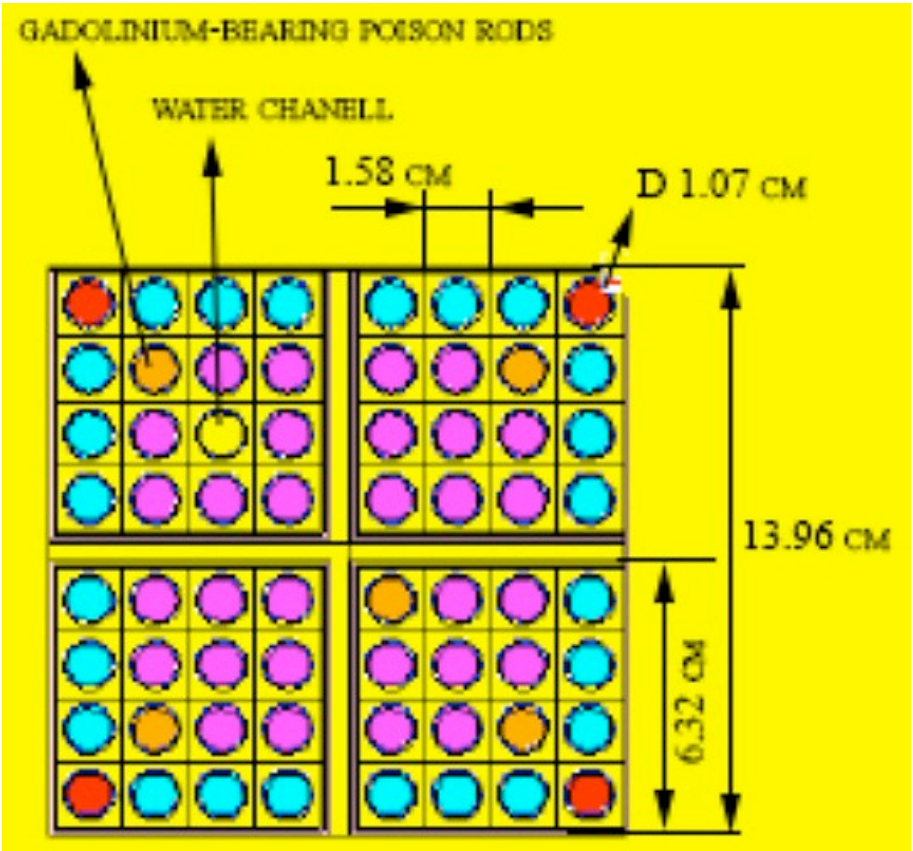

Figure 4: The model and design dimensions of the SVEA-64 assembly.

The setup used for numerical treatment of the Gamma Differential Die-Away approach is shown in Figure 5. It should be noticed that this is not an optimized setup and therefore, further optimization is required. Eight NaI gamma detectors are used in the setup, though other types of gamma detectors can be easily used in the same manner. A point pulsed neutron source with an energy of 2.5 MeV is surrounded by a deuterium moderator in order to decrease the contribution of the 2.223 MeV gamma to the source-related (and fast fission-related) component of the gamma count characteristic. The numerical studies in this paper were performed by using both the MCNPX [9] and the MCNP-PoliMi [10] Monte Carlo codes.

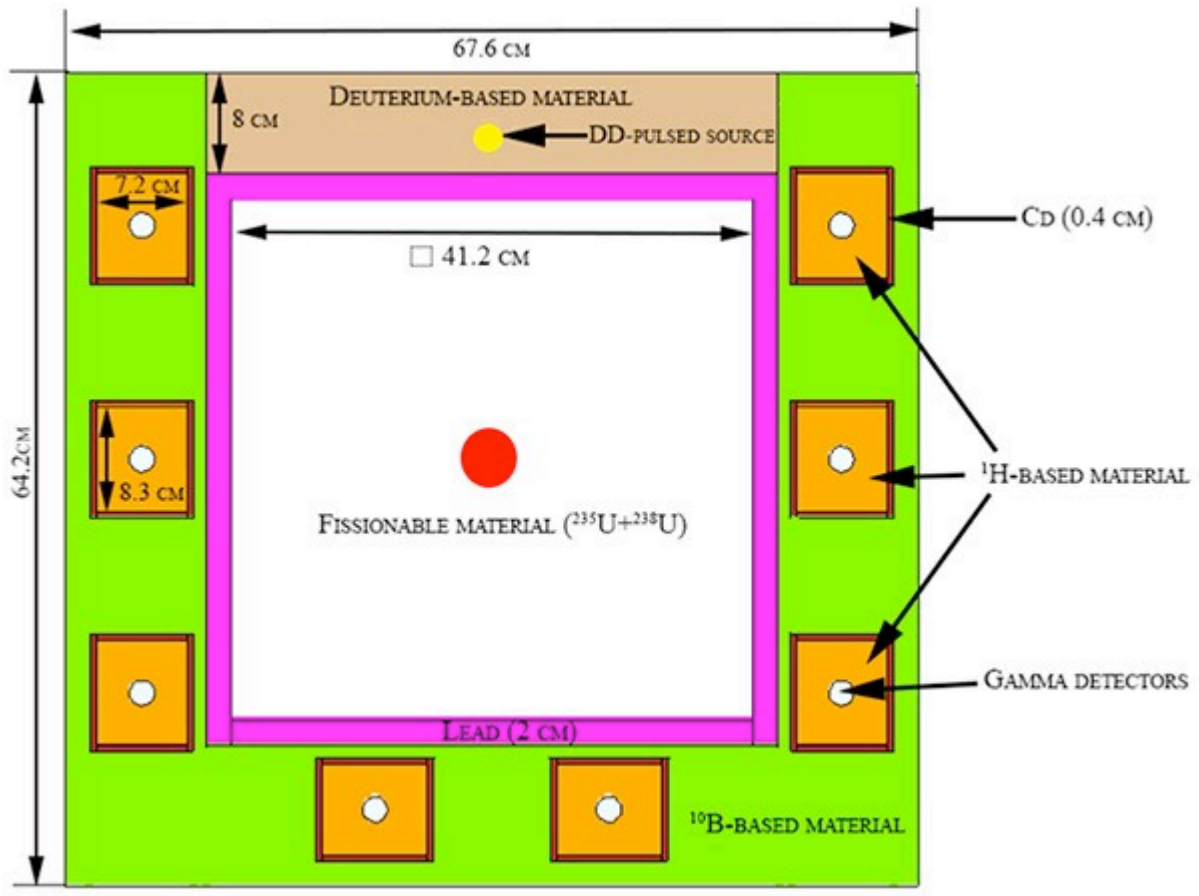

Figure 5: An illustration of the proposed setup for test of Gamma Differential Die-Away approach.

## 4. Results and discussion

### *4.1. A presence of 2.223 MeV line and preliminary observations*

Our previous study of the direct method for evaluating the concentration of a boric acid in a spent fuel pool by using scintillation detectors [11, 12] already indicated the presence of the strong 2.223 MeV gamma line as a result of neutron capture reaction on hydrogen. Therefore, the first step of this work was more or less repetition of a similar experiment but in a different setup and with a NaI gamma detector instead of EJ-309 liquid neutron-gamma scintillation detector. The identification of the 2.223 MeV gamma line in the spectrum was performed via calibration with $^{22}$Na and $^{60}$Co sources which covered the energy range from 0.511 MeV to 2.5 MeV. As shown in Figure 6, the 2.223 MeV gamma line is visibly appearing after just 3 minutes of measurements for both $^{252}$Cf spontaneous fission source of correlated neutrons and an $^{241}$Am-Be ($\alpha$ ,n)-source of random neutrons that are placed in center of the water tank shown in Figure 2. The results of numerical simulations, though, show that most of the 2.223 MeV gammas are produced by the $^{252}$Cf-source in a radius of approximately 10 cm (Figure 7, l.h.s.) with a maximum at the distance of 4 cm in the water (Figure 7, r.h.s.). This observation was used for an arrangement of the experimental setup shown in Figure 2. Numerical simulations indicate as well that the 2.223 MeV gamma produced by a $^{252}$Cf-source in water will have a multiplicity distribution similar to the count distribution shown in Figure 8 (l.h.s.). Considering only two correlated gamma at the time which reach the detector (Figure 8, r.h.s.) and their respective energy deposition in the detector, then most of the correlations will be observed for 1.201 MeV (DE) and 2.223 MeV gamma energy.

It should be noticed that Figure 7 and Figure 8 are results of numerical simulations, should therefore be taken with appropriate caution. Even so we did not observe fundamental disagreement with the results of further experimental studies.

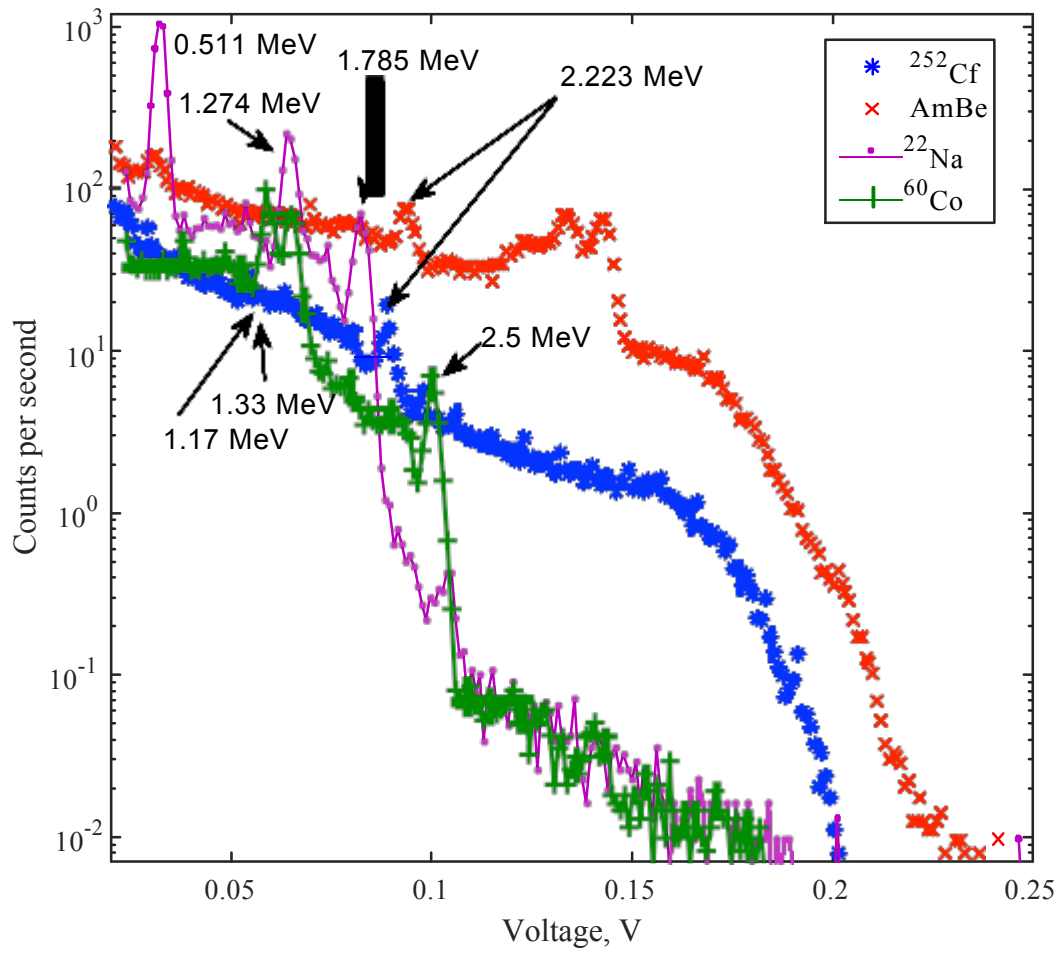

Figure 6: Experimental gamma spectra for $^{252}$Cf, $^{241}$Am-Be, $^{22}$Na and $^{60}$Co sources

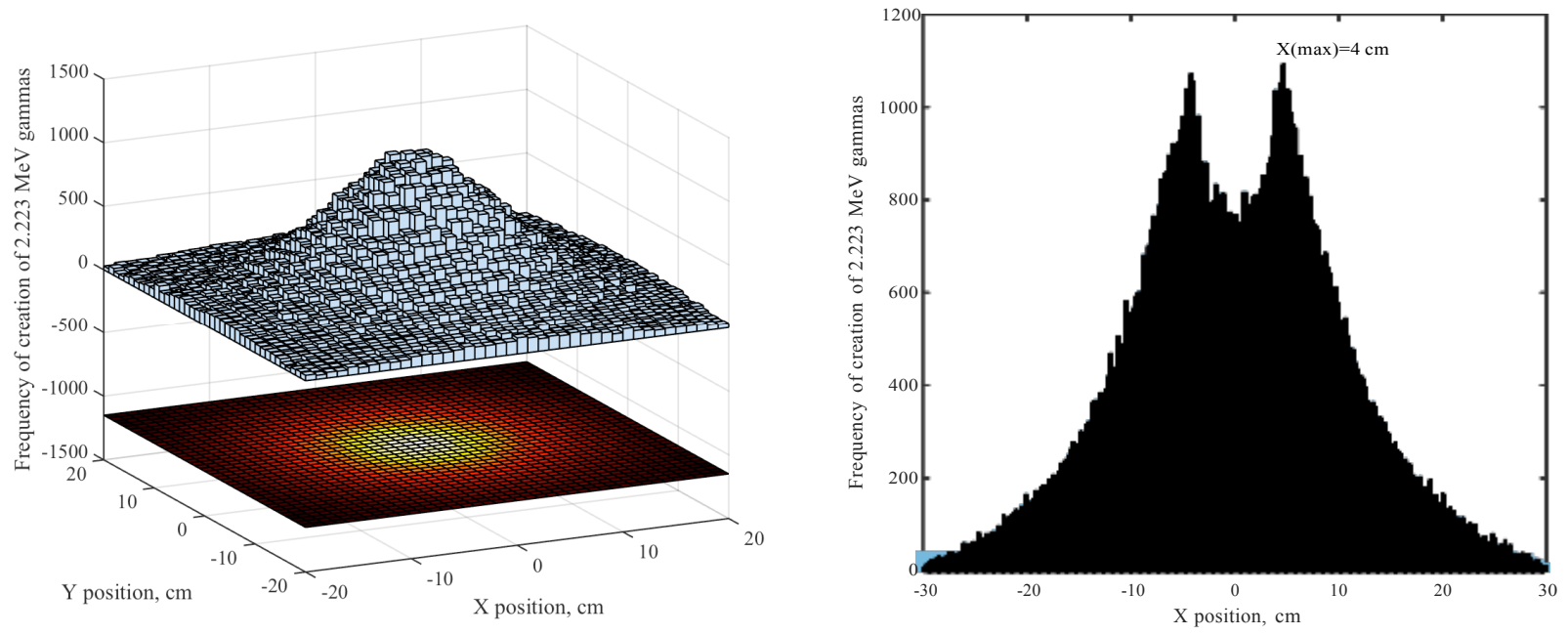

Figure 7: Dependence of the frequency of the 2.223 MeV gamma production by $^{252}$Cf-source in the water on the distance from the source.

### 4.2. *Rossi-alpha and Feynman-alpha*

The next step of preliminary studies was related to numerical investigations of the assumption that "the time scale for correlation of the capture gammas is preserved in the same way as the one for thermal neutrons. Therefore, we made a numerical model of the two sources, $^{240}$Pu spontaneous fission source of correlated neutrons and a 2.5 MeV-source of random neutrons, placed in the center of the water moderator. Thereafter, timestamps of the 2.223 MeV gammas were collected and post processed the same way as described in Section 3.3."Data processing procedure". As expected, results shown in Figure 9 indicate that in the case of only random events being detected (r.h.s.) the Rossi-alpha distribution is constant with time and vice versa for the case of correlated events.

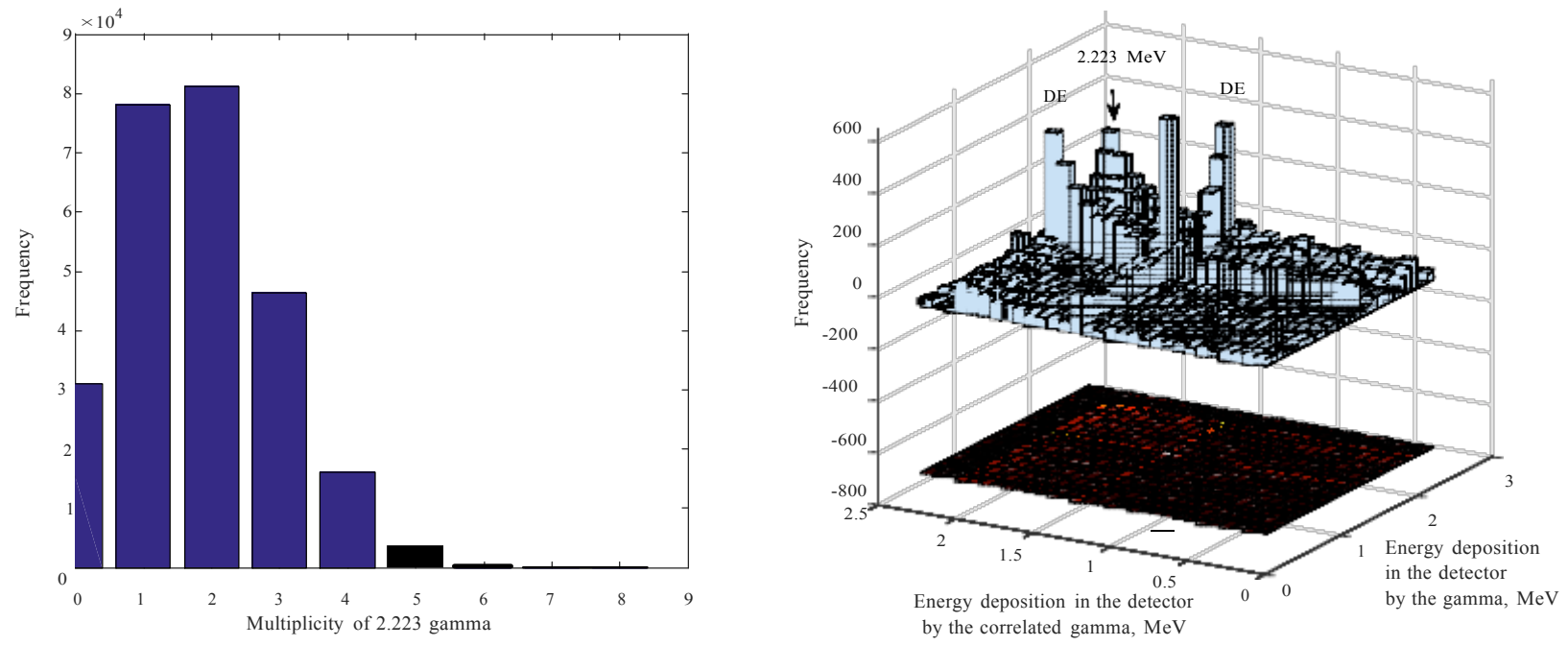

Figure 8: Simulated count distribution for the 2.223 MeV gamma produced by $^{252}$Cf-source in the water (l.h.s.). Frequency of two correlated gamma and their energy deposition in the detector (r.h.s.).

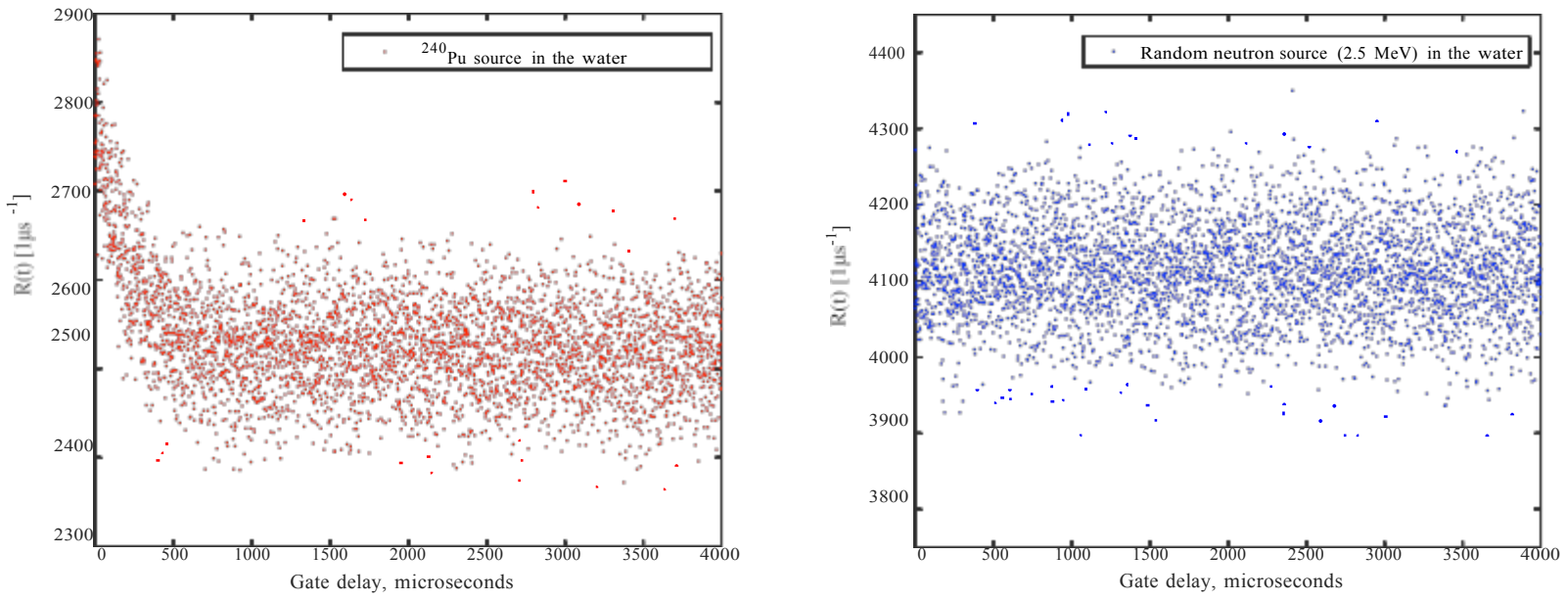

Figure 9: The traditional Rossi-alpha curve for 2.223 MeV gamma produced by the $^{240}$Pu spontaneous fission source of correlated neutrons (l.h.s.) and a 2.5 MeV-source of random neutrons (r.h.s.) in water.

Figure 10 shows results of similar studies, the traditional Rossi-alpha curve for 2.223 MeV gamma but produced by a SVEA-64 spent fuel assembly placed in the spent fuel pool. Thus, there is a positive indication that the same methodology might be used for measurements of the spent fuel directly in a spent fuel pool.

The experimental studies were performed with $^{252}$Cf and $^{241}$Am-Be sources and one NaI detector. Assuming relative dependence between angular correlations of 2.223 MeV gammas [1] and original neutron correlations, and based on the results of the studies of the angular correlations in the prompt neutron emission in spontaneous fission of $^{252}$Cf [13], which investigated neutron-neutron correlations in a plane perpendicular to the fission axis, it was found acceptable if only one detector with a diameter of 8 cm was used in the present experiments.

Since, in the case of the $^{241}$Am-Be source only random events were detected, the

[1] While saying ”2.223 MeV gammas” in the text, in experiment we consider gamma counts in a photo-peak.

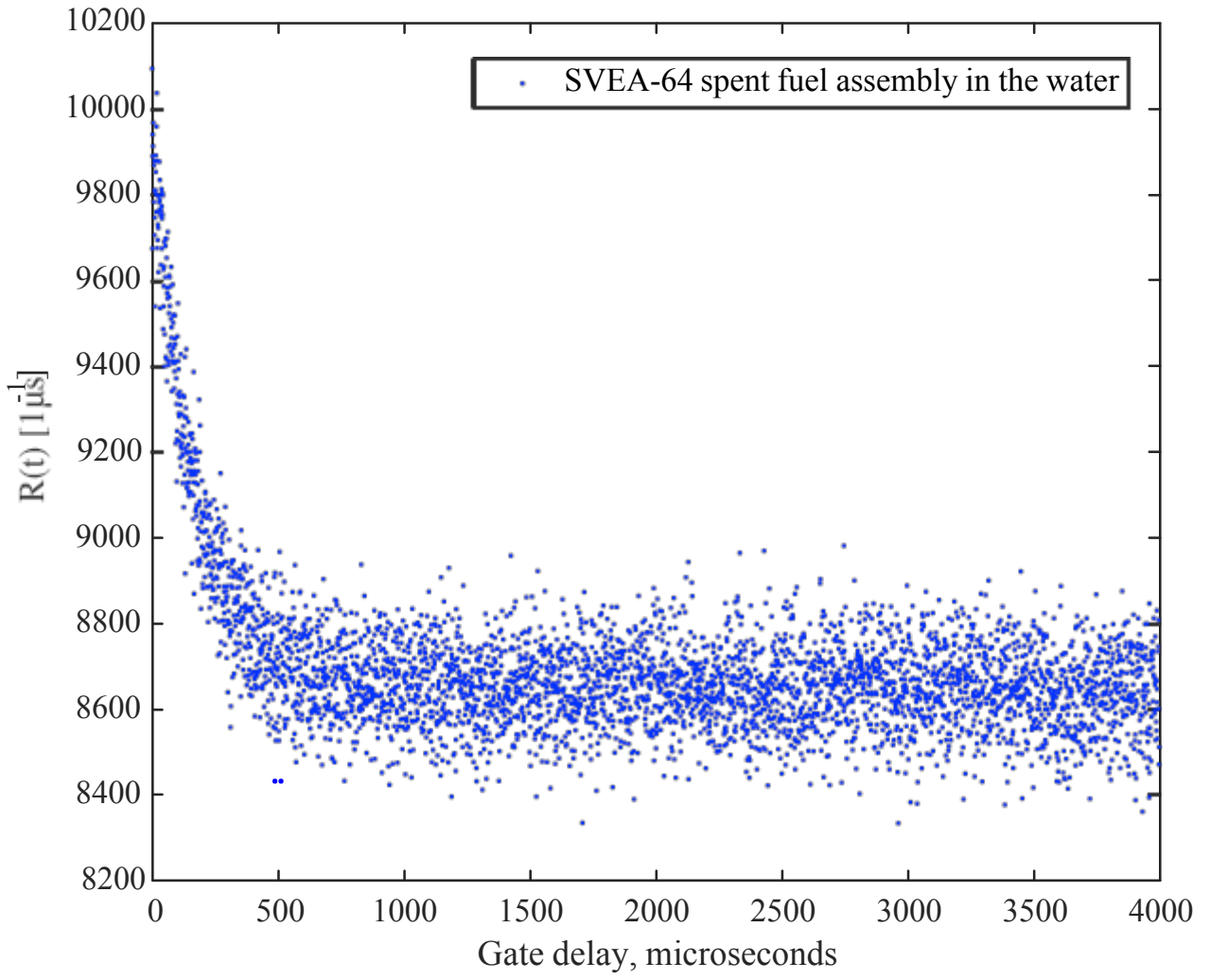

Figure 10: The traditional Rossi-alpha curve for 2.223 MeV gamma produced by a SVEA-64 spent fuel assembly in the spent fuel pool.

Rossi-alpha distribution is constant with time, which is what is shown in both experimental (Figure 11, l.h.s.) and numerical (Figure 11, r.h.s.) investigations.

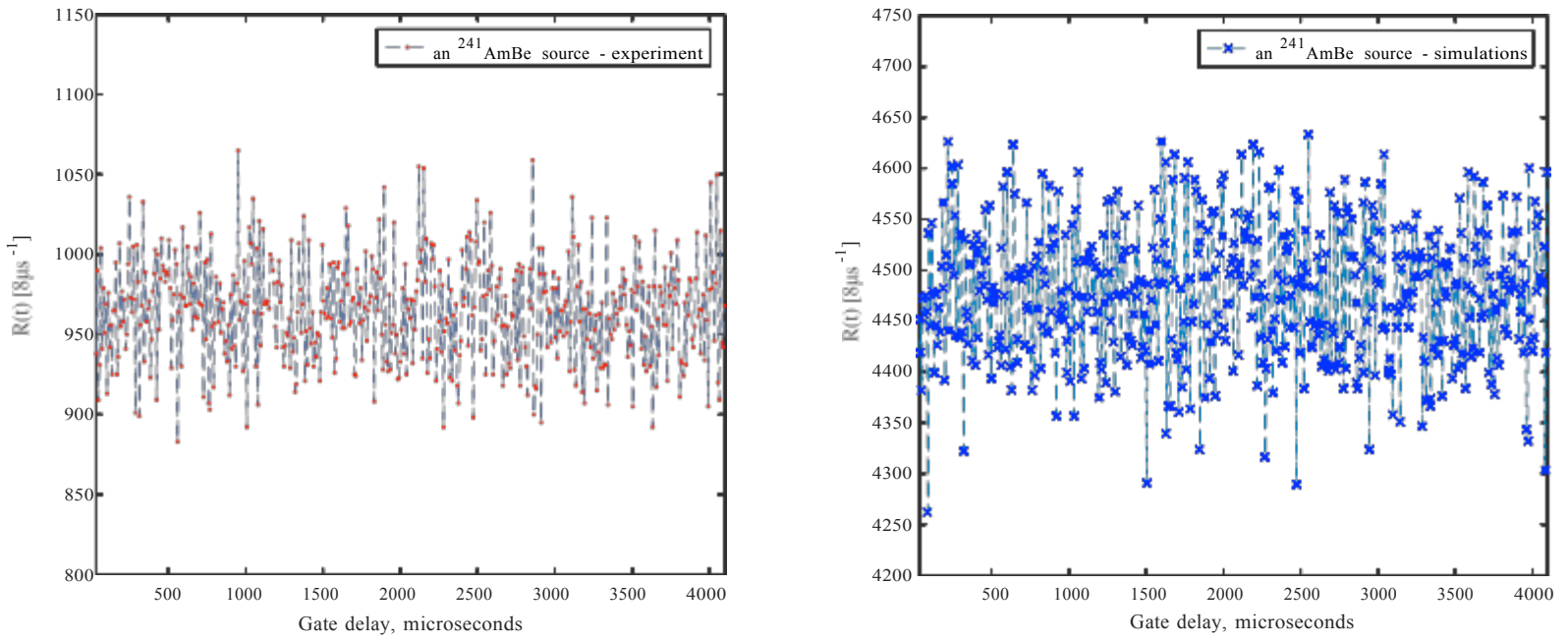

Figure 11: The experimental (l.h.s.) and numerical (r.h.s.) Rossi-alpha curves for 2.223 MeV gammas produced by $^{241}$Am-Be source placed in the water tank (Figure 2).

At the same time the situation is different for the $^{252}$Cf source, as shown in Figure 12, the Rossi-alpha distribution is not constant with time and can be approximated with one exponential function. It should be noted that the $^{252}$Cf source was very weak (17.3 kBq), therefore, measurement time was 2 hours.

As expected, the variance to mean ratio also indicate a non-Poisson nature of particle emission from the $^{252}$Cf source, it deviates from unity for the case of 2.223 MeV gammas produced by a $^{252}$Cf source in water.

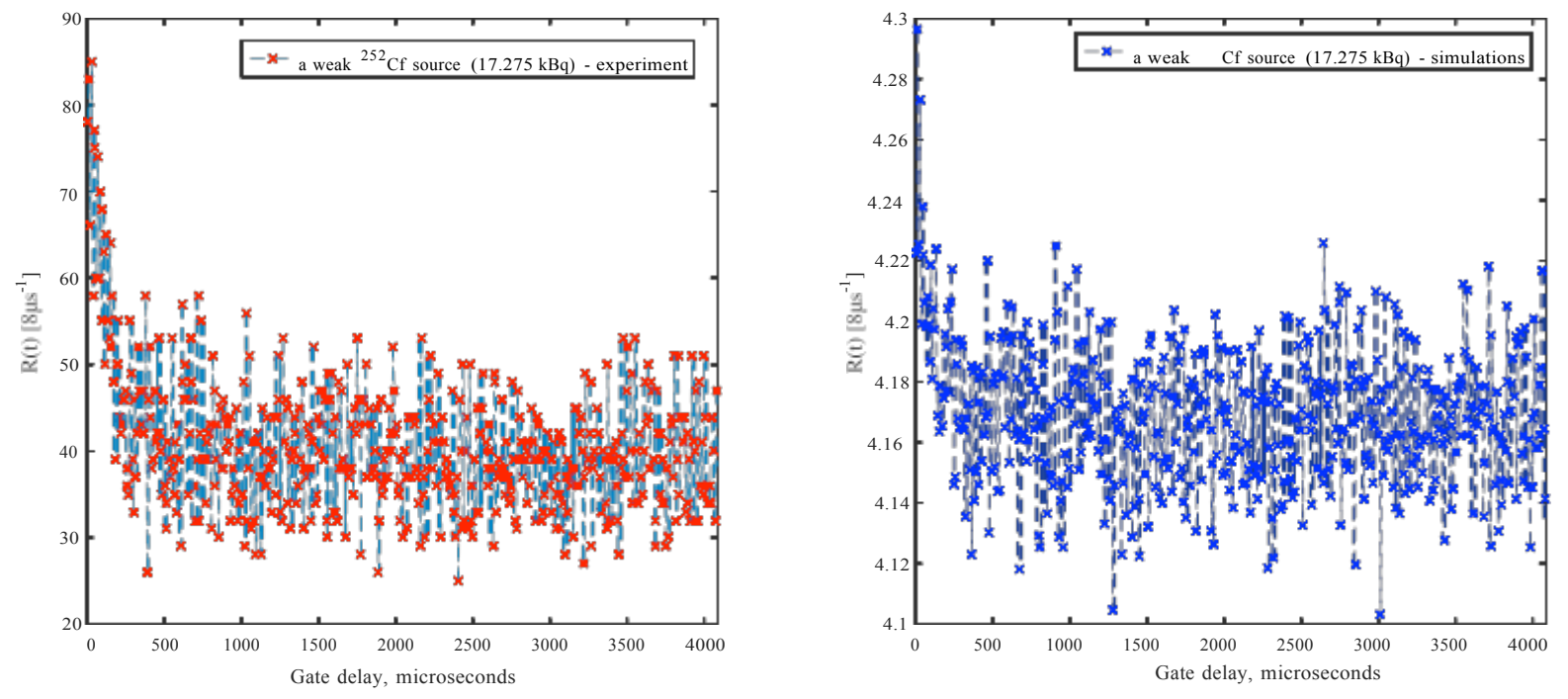

Figure 12: The experimental (l.h.s.) and numerical (r.h.s.) Rossi-alpha curves for 2.223 MeV gammas produced by $^{252}$Cf source placed in the water tank (Figure 2).

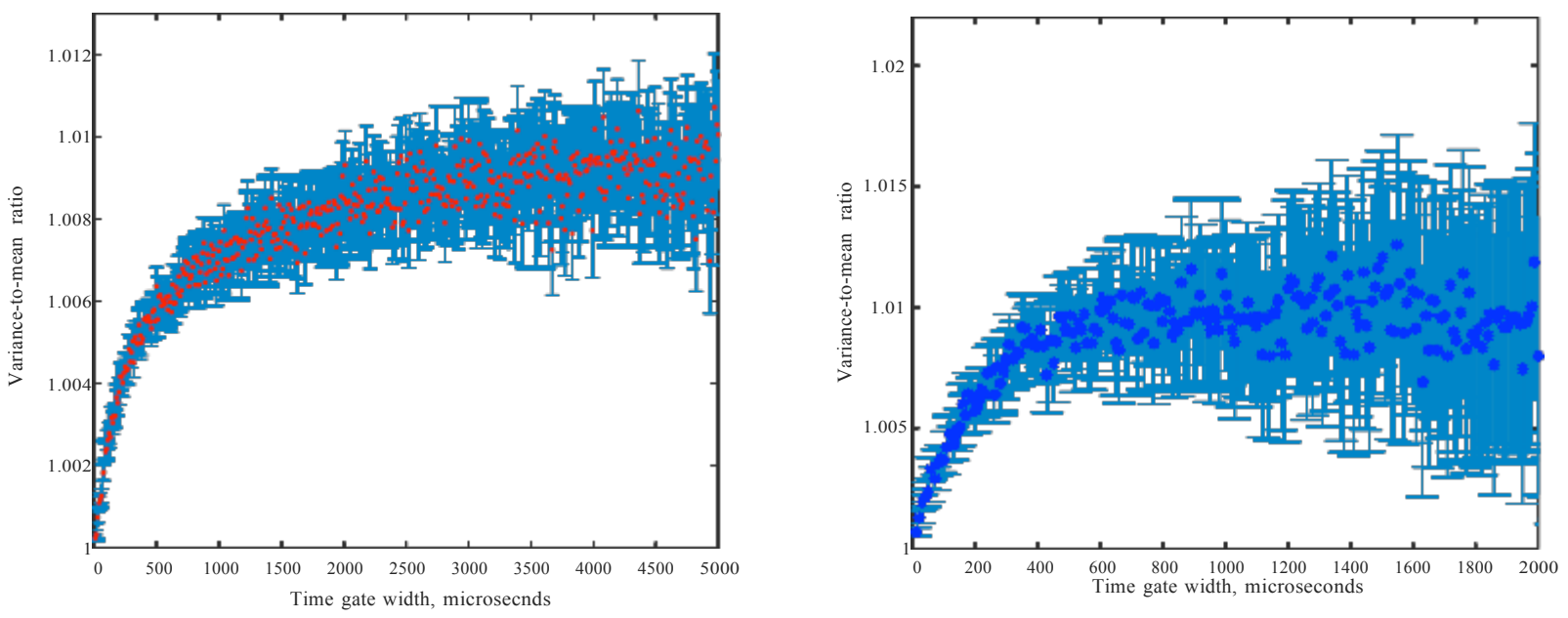

Figure 13: The dependence of the ratio of the variance to mean of the number of 2.223 MeV gamma on the detection time for $^{252}$Cf-source placed in the water tank (Figure 2).

Thus, both experimental and numerical studies show a strong potential for 2.223 MeV gammas (or 1.201 MeV, DE) originated in $^{1}H(n,\gamma)^{2}D$-reaction to complement or be used as a replacement for thermal neutron-based counting. Also, the time scale of correlated 2.223 MeV gammas is preserved in the same way as the one used for correlated thermal neutrons.

## 5. Gamma Differential Die-Away concept

As was shown above, 2.223 MeV gammas from neutron capture reaction on hydrogen behave similarly to thermal neutrons, therefore most of the techniques used in traditional thermal-based counters can be applied to 2.223 MeV gamma correlation counting without the significant modifications. However, the Gamma Differential Die-Away concept which we offer to use as a potential alternative to the traditional Differential Die-Away Analysis [14–16] requires a bit more attention. This is due to the fact that 2.223 MeV gamma originated in $(n,\gamma)$-reaction on the hydrogen, will be created in the system as a whole and detected at anytime after creation by gamma detectors and the Cd screen will obviously not play the same role as in the case of the traditional Differential Die-Away Analysis. Therefore, if the Gamma Differential Die-Away concept is to be realized the outer hydrogen moderator must be replaced with a deuterium-based or boron-based material. Due to its cross-section, see Figure 14 (l.h.s.), in this work boron-based material was eventually used (see Figure 5).

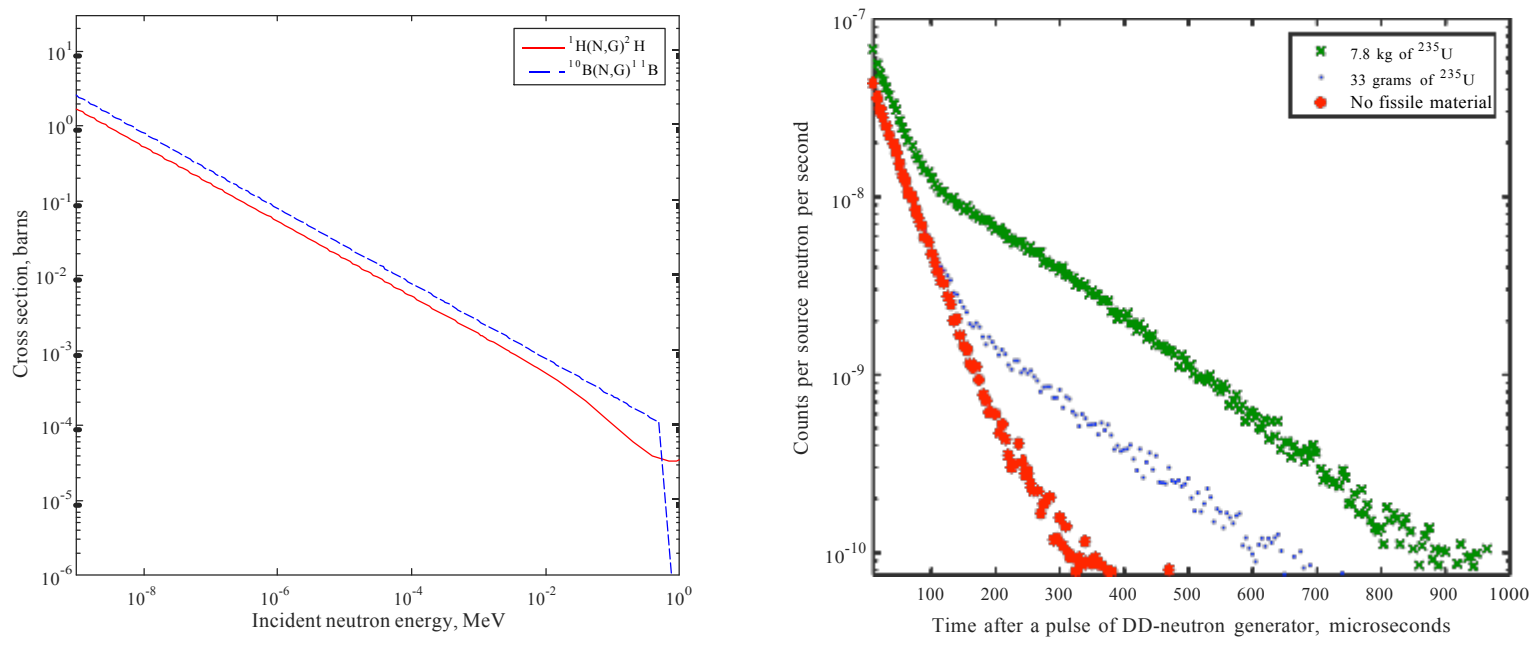

Figure 14: Cross section (ENDF/B-VII.1) [? ] (l.h.s.). Dependence of the number of detections of 2.223 MeV gamma on a time after the pulse of neutron generator (r.h.s.).

Three different cases were considered in order to estimate the performance of the Gamma Differential Die-Away concept, i.e. when only the source was in the system, or 33 grams and 7.8 kilograms of $^{235}U$. Results shown in Figure 14 (r.h.s.) indicate that all these cases can relatively easy be identified. The 2.223 MeV gammas resulting from the capture of source neutrons can be approximated by one exponential, at the same time when $^{235}U$ present in the system, the response can be represented with a two-exponential function. These results might be improved through additional optimization of the thickness and composition of moderators.

## 6. Conclusion

The new concept of using 2.223 MeV gammas (or 1.201 MeV, DE) originating from $^{1}H(n,\gamma)^{2}D$-reactions as an alternative or complement to counting thermal neutrons was evaluated experimentally with a $^{252}$Cf spontaneous fission source of correlated neutrons and an $^{241}$Am-Be ($\alpha$ ,n)-source of random neutrons and numerically with $^{252}$Cf, $^{241}$Am-Be, $^{240}$Pu, a 2.5 MeV random neutron source and a SVEA-64 spent fuel assembly. Both experimental and numerical results indicate that 2.223 MeV gammas behave in a similar manner to the thermal neutrons and at the same microseconds time scale. Moreover, numerical results show that approximately half of the emitted neutrons (for a case of a $^{252}$Cf source in the water tank) lead to the creation of 2.223 MeV gammas. Most of the 2.223 MeV gammas are produced in a radius of approximately 10 cm with a maximum at 4 cm in the water. It was shown that this concept can also be used for the measurements of spent fuel directly in a spent fuel pool. As a part of the work the Gamma Differential Die-Away concept was suggested and tested numerically. It was shown that the presence of $^{235}U$, chosen as an example, can be easily identified. Though, further work is needed to optimize the concept performance. Overall, this proof of concept study indicated that traditional gamma detectors can be used as an alternative to thermal neutron detectors in correlation-based measurements in Nuclear Safeguards and Security.

**Acknowledgement**

This work was supported by the Swedish Radiation Safety Authority, SSM. A research grant from Carl Tryggers Stiftelse made it possible to acquire the measuring equipment. The authors want to thank **Prof. David Wehe**, **Prof. Imre Pázsit, Prof. Martyn T. Swinhoe** and **Andrea Favalli** for useful discussions and advice.

**References**

[1] A.J. Keller, An Increasingly Rare Isotope, Physics 241, Stanford University, 2011.

[2] R.T. Kouzes, The $^{3}$He Supply Problem, Report No. PNNL-18388, Pacific Northweest National Laboratory, 2009.

[3] Current Status of $^{3}$He Alternative Technologies for Nuclear Safeguards, LA-UR-15-21201 Ver. 2, 2015

[4] D. Chernikova, K. Axell, S. Avdic, I. Pázsit, A. Nordlund, The neutron-gamma Feynman variance to mean approach: gamma detection and total neutron-gamma detection, Nuclear Instruments and Methods in Physics Research A, 782 (2015) 4755.

[5] D.Chernikova, K. Axell, A. Nordlund, A theoretical and experimental investigation of using Feynman-Y functions for the total and gamma detections in a nuclear and radioactive material assay, ESARDA (2015).

[6] I. Pázsit, L. Pál, Neutron Fluctuations: A Treatise on the Physics of Branching Processes, Elsevier Science Ltd., London, New York, Tokyo, 2008.

[7] Y. Kitamura, T. Misawa, A. Yamamoto, Y. Yamane, C. Ichihara, H. Nakamura, Feynman-alpha experiment with stationary multiple emission sources, Progress in Nuclear Energy, 48 (2006) 569-577.

[8] B.D. Murphy, I.C. Gauld, Spent Fuel Decay Heat Measurements Performed at the Swedish Central Interim Storage Facility, Report by Oak Ridge National Laboratory, Managed by UT-Battelle, LLC Oak Ridge, TN 37831-6170, NRC Job Code Y6517, 2008

[9] D. B. Pelowitz, MCNPX User Manual, Version 2.7.0, Los Alamos National Laboratory report LA-CP-11-00438, April 2011.

[10] S.A. Pozzi, E. Padovani, M. Marseguerra, MCNP-PoliMi: a Monte-Carlo code for correlation measurements, Nuclear Instruments and Methods in Physics Research Section A, 513 (2003) 550-558.

[11] D. Chernikova, K. Axell, I. Pázsit, A. Nordlund, R. Sarwar, A direct method for evaluating the concentration of boric acid in a fuel pool using scintillation detectors for joint-multiplicity measurements, Nuclear Instruments and Methods in Physics Research Section A, (2013) 90-97.

[12] D. Chernikova, K. Axell, I. Pázsit, A. Nordlund, P. Cartemo, Testing a direct method for evaluating the concentration of boron in a fuel pool using scintillation detectors, and a $^{252}$Cf and an $^{241}$Am-Be source, ESARDA (2013).

[13] Yu. Kopatch, A. Chietera, et al., Detailed Study of the Angular Correlations in the Prompt Neutron Emission in Spontaneous Fission of $^{252}$Cf, Physics Procedia, 64 (2015) 171-176.

[14] J.T. Caldwell, R.D. Hastings, G.C. Herrera, W.E. Kunz, et al., The Los Alamos second-generation system for passive and active neutron assays of drum size report LAB10774 MS (1986).

[15] K.A. Jordan and T. Gozani, Pulsed Neutron Differential Die Away Analysis for Detection of Nuclear Materials, Nuclear Instruments and Methods in Physics Research Section B, 261 (2007) 365-368.

[16] Y. N. Barmakov, E. P. Bogolyubov, O. V. Bochkarev, Y. G. Polkanov, V. L. Romodanov, D. N. Chernikova, System of combined active and passive control of fissile materials and their nuclide composition in nuclear wastes, International Journal of Nuclear Energy Science and Technology, 6(2) (2011) 127-135.